\documentclass[%
 aip,
 amsmath,amssymb,
 reprint,%
]{revtex4-1}

\usepackage{graphicx}
\usepackage{dcolumn}
\usepackage{bm}

\usepackage[utf8]{inputenc}
\usepackage[T1]{fontenc}
\usepackage{mathptmx}
\usepackage{mathrsfs}
\usepackage{braket}
\usepackage{dcolumn}
\usepackage{bm}
\usepackage{tikz}
\usepackage{nicefrac}
\usepackage{array}
\usepackage{graphics}
\usepackage{amssymb}
\usepackage{amsmath}
\usepackage{graphicx}
\usepackage{multirow}
\usepackage{color}
\usepackage{comment}
\usepackage[normalem]{ulem}
\usepackage{url}
\usepackage{mathrsfs}
\usepackage{tabularx}
\usepackage[caption=false]{subfig}
\usepackage{appendix}
\usepackage{float}
\usepackage[makeroom]{cancel}

\usepackage{csquotes}
\renewcommand{\b}[1]{\boldsymbol{#1}}
\newcommand{\nid}{\noindent}

\begin{document}

\preprint{AIP/123-QED}

\title{Dissipative Tunneling Rates through the Incorporation of First-Principles Electronic Friction in Instanton Rate Theory I: Theory}

\author{Y. Litman}
\email{yairlitman@gmail.com}
\affiliation{MPI for the Structure and Dynamics of Matter, Luruper Chaussee 149, 22761 Hamburg, Germany}

\author{E. S. Pós }%
\affiliation{MPI for the Structure and Dynamics of Matter, Luruper Chaussee 149, 22761 Hamburg, Germany}

\author{C. L. Box}%
\affiliation{Department of Chemistry, University of Warwick, Coventry CV4 7AL, United Kingdom}

\author{R. Martinazzo}
\affiliation{Department of Chemistry, Università degli Studi di Milano, Via Golgi 19, 20133 Milano, Italy}

\author{R. J. Maurer}%
\affiliation{Department of Chemistry, University of Warwick, Coventry CV4 7AL, United Kingdom}

\author{M. Rossi}
\email{mariana.rossi@mpsd.mpg.de}
\affiliation{MPI for the Structure and Dynamics of Matter, Luruper Chaussee 149, 22761 Hamburg, Germany}

\date{\today}

\begin{abstract}

Reactions involving adsorbates on metallic surfaces and impurities in bulk metals are ubiquitous in a wide range of technological applications. The theoretical modelling of such reactions presents a formidable challenge for theory because nuclear quantum effects (NQEs) can play a prominent role and the coupling of the atomic motion with the electrons in the metal gives rise to important non-adiabatic effects (NAEs) that alter atomic dynamics. 
In this work, we derive a theoretical framework that captures both NQEs and NAEs and, due to its high efficiency, can be applied to first-principles calculations of reaction rates in  high-dimensional realistic systems. In more detail, we develop a method that we coin  ring polymer instanton with explicit friction (RPI-EF), starting from the ring-polymer instanton formalism applied to a system-bath model. We derive general equations that incorporate the spatial and frequency dependence of the  friction tensor, and then combine this method with the \textit{ab initio} electronic friction formalism for the calculation of thermal reaction rates. We show that the connection between RPI-EF and the form of the electronic friction tensor presented in this work does not require any further approximations, and it is expected to be valid as long as the approximations of both underlying theories remain valid. 
\end{abstract}

\maketitle

\section{\label{sec:intro}Introduction}

Metallic systems lack an energy gap between unoccupied and occupied electronic states. Thus, low energy excitation and deexcitation of electron-hole pairs can easily exchange energy with nuclear vibrations, representing  a violation of the Born-Oppenheimer principle, where electrons are assumed to adjust adiabatically to the position of the nuclei.
This type of non-adiabatic effect (NAE) has been verified experimentally numerous times in the past \cite{Bartels_ChemSci_2011,Wodtke_ChemSocRev_2016,Auerbach_2021} and found to be particularly important for hot-electron-induced reactions \cite{Schindler_JCP_2011}, surface scattering\cite{Cohen_Nat_2005, Box_Jacs_2020} and vibrational relaxation lifetimes \cite{Persson1982,Tully_1993,Rittmeyer_PRL_2015}. Many theoretical approaches have been developed to account for NAEs in these contexts \cite{Dou_JCP_2016,Shenvi_JCP_2009,Ryabinkin_JCPL_2017,Head_Gordon_1995}. Among them, 
the method coined ``molecular dynamics with electronic friction''\cite{Head_Gordon_1995} (MDEF) has the advantage of being a method that can currently be coupled to  \textit{ab initio} electronic structure theory without resorting to model parametrization.~\cite{Maurer_2016_PRB,BlancoRey_PRL_2014} 

MDEF describes the motion of classical nuclei through a  Langevin equation where the friction forces and the corresponding random noise embody the effects of the electronic excitations, therefore approximately including NAEs to an otherwise classical nuclear dynamics. The electronic friction tensor lies at the core of the definition of the friction force in MDEF and can be understood as a first order correction to the Born-Oppenheimer approximation in the presence of a manifold of fast relaxing electronic states~\cite{Dou_JCP_2018}. As a consequence, MDEF is expected to break down when strong non-adiabatic effects are present~\cite{Coffman_PCCP_2018,Bartels_ChemSci_2011} and in cases where charge transfer mechanisms are dominant~\cite{Shenvi_JCP_2009}. Despite these shortcomings, MDEF has proven to be a useful approach for several realistic systems and conditions, in which high-level simulations can explain well-controlled experiments~\cite{Box_Jacs_2020,zhang_hot-electron_2019,spiering_orbital-dependent_2019,Kandratsenka_PNAS_2018,Bunermann_Sci_2015,Rittmeyer_PRL_2015}. 

The classical-nuclei approximation in the context of MDEF is appropriate for many situations. However, when studying the motion of light atoms, such as hydrogen, deuterium, lithium, etc., the quantum nature of the nuclei can lead to significant quantum effects including tunneling, isotope effects, and non-Arrhenius temperature dependence of rates~\cite{Makri_JCP_1998,Kimizuka_PRB_2019,RossiMano2016,Kimizuka_PRM_2021}. Indeed, Fang \textit{et al}. \cite{Fang_PRL_2017} showed that depending on the shape of the barrier for hydrogen diffusion on metallic surfaces, a coexistence of deep tunneling through the barrier and classical hopping above the barrier can take place even at elevated temperatures, while Kimizuka \textit{et al}. exposed a strong interplay between the relevance of the quantum fluctuations and lattice strain of interstitial H diffusion in metals \cite{Kimizuka_PRB_2018}. 
 
 In this work, we propose a new approach based on ring polymer instanton (semi-classical) rate theory \cite{Jor_review_2018},  that includes NAEs through the electronic friction formalism initially proposed by
 Hellsing and Persson \cite{Hellsing_1984} and 
 Head-Gordon and Tully\cite{Head_Gordon_1995}. As we will  discuss later, the description of the electrons as a harmonic bath of non-interacting particles is the key consideration that makes this connection possible. We show how both theories can be combined naturally and include the full frequency and position dependence of the electronic friction. 
 
 Part I of this paper is organized as follows: In section \ref{sec:RPI} we present a brief review of the ring-polymer instanton (RPI) theory. In section \ref{sec:RPI-EF} we introduce and discuss our proposed theory coined ring polymer instanton with explicit friction (RPI-EF). In section \ref{sec:aifric} we present and discuss the connection of RPI-EF and an \textit{ab initio} electronic friction, and finally conclude part I of this paper in section \ref{sec:Conclusions}. In part II, we present benchmarks of our method for model systems and an application to hydrogen hopping in Pd, employing Kohn-Sham density-functional theory.

\section{Ring Polymer Instanton Rate Theory \label{sec:RPI}}

The RPI approximation~\cite{JOR_JCP_2009, Arni_thesis} 
 is a semi-classical method based on the path integral formulation of quantum mechanics and allows the evaluation of reaction rates in the deep tunneling regime~\cite{Miller_JCP_1975,Jor_review_2018}. The theory assumes that only a couple of well-defined reactant and product states are sufficient to describe the reactive process under consideration. This condition is met for gas-phase reactions and atomic diffusion on (or in) solids,~\cite{Fang_PRL_2017} but it is rarely satisfied for liquid environments \cite{Grifoni_PNAS_2019}. The RPI approximation replaces the 
quantum mechanical time propagator in the flux-side time correlation function by its semi-classical counterpart, and allows one to express the reaction rate in terms of the dominant stationary trajectory in imaginary time that connects reactants and products, i.e. the  instanton trajectory. This special trajectory can be interpreted as the main tunneling pathway at a given temperature and offers an intuitive picture of the process under study. RPI rate theory has been derived by Richardson  
from quantum scattering theory \cite{Richardson_JCP_2016}
and has been shown by Althorpe to be equivalent to the  more traditional derivation based on the ``Im \textit{F}'' premise\cite{Althorpe_JCP_2011}, where the  rate is related to the imaginary part of the system free energy  \cite{ Affleck_1981_PRL}.

In order to find the instanton pathway, it is numerically convenient to discretize the closed trajectories and represent them by ring polymers  \cite{Jor_review_2018}. The discretized Euclidean action for a given trajectory in imaginary time, $S_P$, is  related to the potential energy of the ring polymer, $U_P$, by
\begin{equation}
      \label{eq:S_P}
   S_P/\hbar = \beta_P U_P ,
\end{equation}
\nid with
\begin{equation}
      \label{eq:U_P}
  U_P (\bm{q}) =\\
 \sum_{k=1}^P 
 \sum_{i=1}^{3N} 
   m_i\frac{\omega_P^2}{2} (q_i^{(k)}-q_i^{(k+1)})^2  +  \\ 
     \sum_{k=1}^P V( q_1^{(k)},\dots q_{3N}^{(k)}).
\end{equation}
\nid Here, $q^{(k)}_i$ is the position of the $i$-th degree of freedom of the $k$-th replica, $m_i$ is the mass of the  $i$-th degree of freedom, $N$ is the number of atoms, $P$ is the number of replicas,
$\bm{q}$ is an abbreviated notation to represent all the degrees of freedom,
and $\omega_P = (\beta_P \hbar)^{-1}$ with $\beta_P = \frac{1}{k_BPT}$, where
 $k_\text{B}$ is the Boltzmann constant and $T$ is the temperature.
 As a result of Eq.~\ref{eq:S_P} and the fact that the instanton pathway is a stationary trajectory in imaginary time, the instanton geometry represents a stationary point on $U_P$. Moreover, it constitutes a first order saddle point and can easily be  found by standard saddle-point search algorithms\cite{Rommel_JCTC_2011}.

The tunneling rate can be expressed as \cite{Jor_review_2018}
\begin{equation}
\begin{split}
   \label{eq:Kinst0}
   k_\text{inst}(\beta) =
   -\frac{2}{\beta\hbar}\text{Im}\hspace{1pt} F =
   \frac{1}{Z^r_P(\beta)}\frac{2}{\beta\hbar}\text{Im}
\int e^{- S_P(\b{q})/\hbar}  d \b{q},
   \end{split}
\end{equation}
\nid where
$F$ is the system's complex free energy and $Z^r_P$ the reactant canonical partition function. The evaluation
of the integral is performed by a steepest-descent integration around the instanton geometry $\b{\bar{q}}$ for all modes with positive eigenvalues, while the mode with negative eigenvalue and the mode with zero eigenvalue, which corresponds cyclic permutation of beads, require special care. As a result, the instanton rate reads
\begin{equation}
\begin{split}
   \label{eq:Kinst1}
   k_\text{inst}(\beta) = 
   \frac{1}{Z_r(\beta)}
    \frac{1}{\beta_P \hbar}
\sqrt{ \frac{B_P(\b{\bar{q}})}{2 \pi \beta_P \hbar^2}}
    Z_\text{vib} 
    e^{-S_P(\b{\bar{q}})/\hbar} ,
       \end{split}
\end{equation}

\nid with
\begin{equation}
\begin{split}
   \label{eq:BP}
   B_P (\b{\bar{q}})= \sum_{i=1}^{3N} \sum_{k=1}^P m_i(\b{\bar{q}}^{(k+1)}_i-\b{\bar{q}}^{(k)}_i)^2,
\end{split}
\end{equation}
\nid and
\begin{equation}
\begin{split}
   \label{eq:Qvib}
Z_\text{vib} = \prod_k {'} \frac{1}{\beta_P\hbar|\lambda_k|}.
\end{split}
\end{equation}

In the expression above, \nid $\lambda_k$ represent the $P\times3N$ eigenvalues of the mass scaled ring-polymer Hessian defined by the second derivatives of $U_P$ with respect to the replica positions, and the prime indicates that the product is taken over all modes except the ones with zero eigenvalue. The contribution of translational and rotational degrees of freedom
have been discarded in Eq.~\ref{eq:Kinst1} since they are not relevant for the present work.
The accuracy of the tunneling rates defined by this theory is limited mainly for two reasons: i) the  fluctuations orthogonal to the reactive direction are considered to be harmonic and, ii) due to lack of real-time information, recrossing effects are completely neglected. 
Despite these shortcomings, the RPI approximation constitutes a valuable and practical method due to its favorable trade-off between accuracy and computational cost, and the method has been successfully applied to systems containing up to several hundreds of degrees of freedom, using an \textit{ab initio} description of their electronic structure  \cite{Rommel_JPCB_2012,Litman_JACS,Litman_PRL}.

Instanton trajectories only exist below  a critical temperature known as cross-over temperature, $ T_c^\circ$, which in most cases can be estimated by  a parabolic barrier approximation as
  \begin{equation}
      \label{eq:Tc}
 k_BT_c^\circ = \frac{\hbar \omega^\ddag}{2\pi},
\end{equation}
 where $\omega^\ddag$ represents the imaginary frequency at the barrier top between reactants and products. At temperatures  below $T_c^\circ$ the reactive process is dominated by tunneling, 
 while above $T_c^\circ$ the classical `over-the-barrier hopping' mechanism represents the major contribution, with nuclear tunneling playing a minor role \cite{Gillan1987}.
 Due to the lack of real-time information the RPI approach is sometimes presented as a `thermodynamic' method \cite{Weiss_book}, and therefore can be seen as an extension of the Eyring transition state theory into the deep tunneling regime (i.e.  extension for temperatures below $T_c^\circ$).

\section{Ring Polymer Instanton Rate Theory with Explicit Friction} \label{sec:RPI-EF}

We consider a system coupled to a harmonic bath, 
which leads to the following modified RP potential
\begin{equation}
\begin{split}\label{eq:UP_sb}
U_P^\text{sb} &= U_P^\text{sys} + 
\sum_{j=1}^{N_b}  \bigg[ \sum_{k=1}^{P}
\frac{\mu_j\omega_P^2}{2} (x_j^{(k)}-x_j^{(k+1)})^2 +
\\
&
\sum_{i=1}^{3N} \sum_{k=1}^{P} \frac{\mu_j\omega_j^2}{2}\left( x_j^{(k)} - \frac{f_{ij}(\b{q}^{(k)})}{ \mu_j \omega_j^2}\right)^2
\bigg],
\end{split}
\end{equation}
\nid where 
$U_P^\text{sys}$ is the system RP potential given by Eq.~\ref{eq:U_P}, $\b{q}^{(k)}=\{q_1^{(k)},q_2^{(k)},\dots, q_N^{(k)}\}$,
$N_b$ is the total number of bath modes,
$x_j^{(k)}$ represents the coordinate of the $k$-th replica of the  $j$-th bath mode, 
 and $\mu_j$ and $\omega_j$ are its corresponding mass and frequency, respectively.
The function $f_{ij}$ determines the coupling between the $j$-th bath mode and the  $i$-th system degree of freedom, even though it is, in principle, a function of all the system degrees of freedom \cite{CALDEIRA_1983}.

The harmonic bath can be completely characterized by a second-rank tensor known as spectral density  whose components  are given by 
\begin{equation}
\begin{split}\label{eq:J}
J_{il}(\b{q},\omega) =
\frac{\pi}{2}\sum_{j=1}^{N_b} 
\left(\frac{\partial f_{ij}(\b{q})}{\partial q_l}\right)^2
\frac{1}{\mu_j\omega} (\delta(\omega-\omega_j)+\delta(\omega+\omega_j)),
\end{split}
\end{equation}
\nid and the time- and position-dependent friction tensor, which will be an important quantity in this paper, can be expressed in terms of the spectral density as
\begin{equation}
\begin{split}\label{eq:eta_t}
\eta_{il}(\b{q},t) = \frac{1}{\pi} \int_{-\infty}^{\infty}d\omega \frac{J_{il}(\b{q},\omega)}{\omega}\cos(\omega t),
\end{split}
\end{equation}
\nid where it is understood that the previous equation is valid only for $t\geq0$ \cite{Weiss_book}.

For reasons that will become clear later, we write the Laplace transform of $\eta_{il}(t)$ as 
\begin{equation}
\begin{split}\label{eq:eta_laplace}
\tilde{\eta}_{il}(\bm{q}, \lambda) =\int_0^{\infty} e^{-\lambda t}\eta_{il}(\bm{q}, t)dt =  \frac{1}{\pi} \int_{-\infty}^{\infty}d\omega \frac{J_{il}(\bm{q},\omega)}{\omega}\frac{\lambda}{\omega^2+\lambda^2}.
\end{split}
\end{equation}

The derivation of the mean-field expression for the ring-polymer potential energy involves a coordinate transformation from the Cartesian representation to the normal modes of the free RP  and a later Gaussian integral as detailed below. 
The quantum canonical partition function, $Z$, for such a system can be related to a classical partition function, $Z_P$, as \cite{Feynman}

\begin{equation}
\begin{split}
Z = \lim_{P\to\infty} Z_P
\end{split}
\end{equation}

\nid with 
\begin{equation}
\begin{split}\label{eq:Qp}
Z_P=& \bigg[\left(\frac{m}{2\pi\beta_P\hbar^2}\right)^{P/2}
\prod_j^{N_b}\left(\frac{\mu_j}{2\pi\beta_P\hbar^2}\right)^{P/2}\bigg]\times\\&
\int dq^{(1)} \dots dq^{(P)}
\prod_j^{N_b} dx_j^{(1)} \dots dx_j^{(P)} e^{- \beta_P U_P^{\text{sb}}}
\end{split}
\end{equation}

\nid where 
$U_P^{\text{sb}}$ is the RP potential of Eq.~\ref{eq:UP_sb}.

We now perform a unitary transformation into the RP normal modes space \cite{Markland_JCP_2008} to get

\begin{equation}
\begin{split}
U_P^{\text{sb}} =&  
\sum_{l=-P/2+1}^{P/2}
\sum_{i=1}^{3N}\frac{1}{2}m_i \omega_l^2(Q_i^{(l)})^2 +
\sum_{k=1}^{P} V( q_1^{(k)},\dots q_{3N}^{(k)}) 
\\&+
\sum_{j=1}^{N_b} 
\sum_{l=-P/2+1}^{P/2} 
\bigg[
\frac{1}{2}\mu_j\omega_l^2(X_j^{(l)})^2+
\sum_{i=1}^{3N}
\frac{1}{2}\mu_j \omega_j^2 \bigg( X^{(l)}_j
-\frac{f_{ij}^{(l)}}{\mu_j\omega_j^2}\bigg)^2
\bigg],
\end{split}
\end{equation}

\noindent where  $X_j^{(l)}$ and $Q_i^{(l)}$ represent coordinates in the normal mode space and $f_{ij}^{(l)}=\sum_{k=1}^P C_{lk}f_{ij}(\textbf{q}^{(k)})$  represents the RP transformed system-bath coupling with $\bm{C}$ being the RP normal mode transformation matrix (see Appendix A) and $\omega_l=2\omega_P \sin(|l|\pi/P)$ . In the previous expression, an even number of replicas, $P$, has been assumed. It is straightforward to treat an odd number of beads, but more involved and not necessary for the present derivation.


In order to perform an integration over the bath degrees of freedom, it is convenient to rewrite the previous equation as

\begin{equation}
\begin{split}
U_P^{\text{sb}} =&  
  \sum_{l=-P/2+1}^{P/2} \sum_{i=1}^{3N}\frac{1}{2}m_i \omega_l^2(Q_i^{(l)})^2 + \sum_{k=1}^{P} V( q_1^{(k)},\dots q_{3N}^{(k)}) 
\\&+
\sum_{l=-P/2+1}^{P/2}  \sum_{j=1}^{N_b} \sum_{i=1}^{3N} \frac{1}{2}\frac{(f_{ij}^{(l)})^2\omega_l^2}{\mu_j(\omega_j^2+\omega_l^2)}
\\&+
\sum_{l=-P/2+1}^{P/2}  \sum_{j=1}^{N_b}  \sum_{i=1}^{3N}
\frac{1}{2}\mu_j (\omega_j^2+\omega_l^2) \bigg( X^{(l)}_j
-\frac{f_{ij}^{(l)}}{\mu_j(\omega_j^2+\omega_l^2)}\bigg)^2.
\end{split}\label{eq:rewrite_UP}
\end{equation}

Introducing Eq.~\ref{eq:rewrite_UP} into Eq.~\ref{eq:Qp} and integrating over the bath degrees of freedom yields 

\begin{equation}
\begin{split}
Z_P= Z_P^\text{bath}\left(\frac{m}{2\pi\beta_P\hbar^2}\right)^{P/2}
\int dq^{(1)} \dots dq^{(P)} e^{- \beta_P U^\text{MF}_P},
\end{split}
\end{equation}

\nid where
\begin{equation}
\begin{split}
Z_P^\text{bath}=& 
\prod_j^{N_b}\prod_{l=-P/2+1}^{P/2}\frac{1}{\beta_P \hbar \sqrt{(\omega_j^2+\omega_l^2)}}
\end{split}
\end{equation}
\nid which converges to the harmonic oscillator partition function $\prod_j^N (2\sinh(\beta \omega_j/2))^{-1}$ in the limit of  $P\to \infty$ \cite{Kleinert}, and the mean-field (MF) RP potential is given by

\begin{equation}
\begin{split}\label{eq:eff}
U^{\text{MF}}_P =&
   U^{\text{sys}}_P
+\sum_{l=-P/2+1}^{P/2} \sum_{i=1}^{3N} \sum_{j=1}^{N}  \frac{1}{2}\frac{\omega_l^2}{\mu_j(\omega_j^2+\omega_l^2)\omega_j^2}(\tilde{f}_{ij}^{(l)})^2.
\end{split}
\end{equation}

From this point, an expression of the rate can be obtained in analogy to Eq.~\ref{eq:Kinst1}.
  We shall refer to this formulation as  RPI with explicit friction (RPI-EF) for reasons that will become clear later. The discretized ring polymer formulation that we present in this paper is theoretically equivalent to previous formulations proposed by Caldeira, Leggett and others~\cite{CALDEIRA_1983,Weiss_book}, but it presents several advantages: it allows for a more intuitive analysis, it is mathematically simpler, and it is computationally more efficient.
We note that related methodologies
have been proposed in the literature before~\cite{Srinath_JCP_2020,Lawrence_JCP_2019,Lawrence_JCP_2020}.

Eq. \ref{eq:eff} nicely shows how, according to quantum mechanics, the effect of the bath modifies  time-independent 
 equilibrium properties, while in the classical limit (i.e. $P=1$, and $\omega_{l=0}=0$) the bath contribution to these properties becomes zero. The MF approximation does not account for dynamical effects of the bath on the system, which makes it particularly suitable to be combined with the ring-polymer instanton method. We note also that even though the random force is a crucial element in the MDEF approach \cite{Hertl_JCP_2021}, rooted in the second fluctuation dissipation theorem \cite{Kubo_1966}, it does not appear in the RPI-EF theory due to the lack of real-time trajectories. Next, we consider different possible forms of the system-bath coupling. 

\subsection{Position Independent Friction}

The first type of coupling considered is a linear coupling given by 
\begin{equation}
\begin{split}\label{eq:f_SI}
f_{ij}(\bm{q})= c_j q_i.
\end{split}
\end{equation}
\nid Using this expression, we can write Eq.~\ref{eq:eff} as
\begin{equation}
\begin{split}\label{eq:MF-SI}
U^{\text{MF}}_P =&   U^{\text{sys}}_P
+\sum_{l=-P/2+1}^{P/2}  \sum_{i=1}^{3N} \bigg[ \sum_{j=1}^{N_b}  \frac{1}{2}\frac{\omega_k^2c_j^2}{\mu_j(\omega_j^2+\omega_k^2)\omega_j^2}\bigg](Q_i^{(l)})^2\\
=& U^{\text{sys}}_P
+
\sum_{l=-P/2+1}^{P/2}   \sum_{i=1}^{3N}\bigg[   \frac{1}{\pi}\int_{-\infty}^{\infty}d\omega\frac{J_{ii}(\omega)}{\omega}\frac{\omega_l^2}{\omega^2+\omega_l^2}\bigg](Q_i^{(l)})^2
\\=&
 U^{\text{sys}}_P +
\sum_{l=-P/2+1}^{P/2}  \sum_{i=1}^{3N}  \frac{\tilde{\eta}_{ii}(\omega_l){\omega_l}}{2}(Q_i^{(l)})^2,\\
\end{split}
\end{equation}
where we used Eq.~\ref{eq:J} in the second line, and Eq.~\ref{eq:eta_laplace}, and Eq.~\ref{eq:f_SI}  in the last line.
Importantly, as a consequence of Eq.~\ref{eq:J}, a linear coupling function results in a friction tensor that is  position independent.

\begin{figure}[h]
\centering
\subfloat[\label{fig:RP_A}]{\includegraphics[width=0.45\columnwidth]{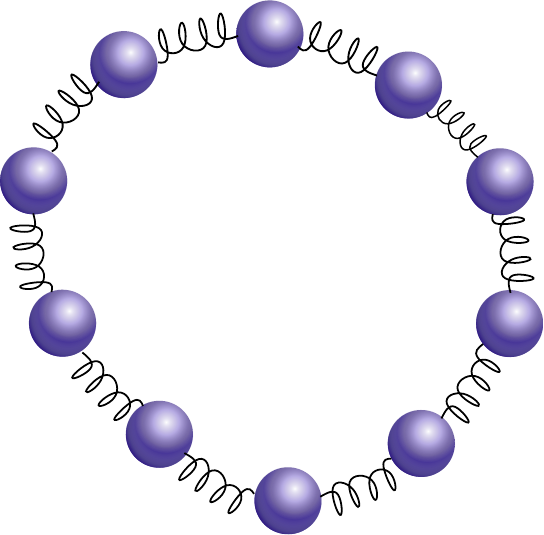}}\hfill
\subfloat[\label{fig:RP_B}] {\includegraphics[width=0.45\columnwidth]{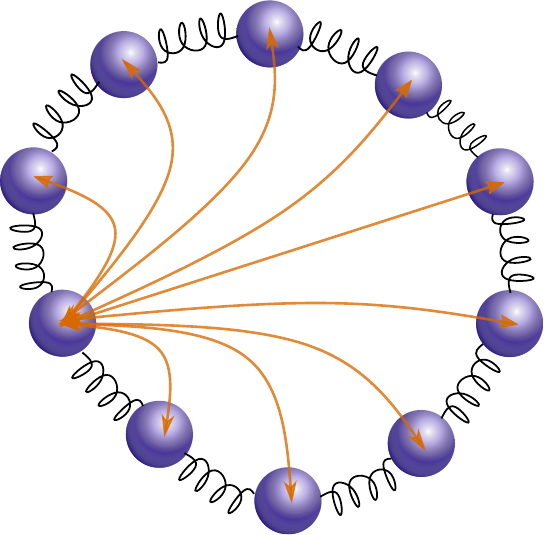}}\hfill
\caption{Cartoon representing the ring polymer of (a) a free particle, and (b) a particle in contact with a harmonic bath. The standard harmonic springs are depicted in black while the springs emerging from the system-bath coupling are depicted in orange. These two couplings correspond to the second and last term on the right hand side of Eq.~\ref{eq:MF-SI}, respectively. Note that, for the sake of clarity, we only draw the latter term for a single bead, but it is present between all pairs of beads.
}
\label{fig:RP}
\end{figure}


In Fig. \ref{fig:RP} we show a cartoon representation of  
the ring polymer corresponding to one free particle and to the same particle coupled to a harmonic bath. The second term on the right-hand side of  Eq.~\ref{eq:MF-SI}, the `friction springs', couple the beads beyond the nearest neighbours, representing a term that is non-local in imaginary time (see further discussion in Appendix B). For a positive-definite friction tensor, these friction spring-terms increase the effective coupling among the beads causing the system to behave more classically when compared to the free-particle system. Indeed, in the low friction limit, the zero-point energy (ZPE) of a damped harmonic oscillator with frequency $\omega$, $E^{\omega,\eta}_\text{ZPE}$, decreases as 
\begin{equation}
\begin{split}
E^{\omega,\eta}_\text{ZPE} -
E^{\omega,0}_\text{ZPE}
\approx - \frac{\hbar \tilde{\eta}_0 \omega_D}{4\omega}
+\frac{\hbar \tilde{\eta}(0) }{2 \pi}
 \ln\frac{\omega_D}{\omega} < 0 ,
\end{split}\label{eq:ZPE}
\end{equation}
\nid where a spectral density with Drude cutoff $J^\text{Drude}(\omega)= \eta\omega \omega_D^2/(\omega_D^2+\omega^2)$ was assumed in the derivation \cite{Ingold_2002}.

Next, we show that when the friction is position-independent, we can derive an extension of the Grote-Hynes approximation for the reaction rate in the deep-tunneling regime.

\subsection*{Extension of the Grote-Hynes approximation into the Deep Tunneling Regime}

Grote, Hynes\cite{Grote_JCP_1980,Grote_JCP_1981} and Pollak \cite{Pollak_JCP_1986}   showed that for intermediate to strong friction values,
the classical reaction rate of the system-bath model can be written in terms of the system rate and $\tilde{\eta}(\omega)$ as 

\begin{equation}
\begin{split}\label{eq:GH}
k_\text{TST}(\tilde{\eta}) = 
k^\text{sys}_\text{TST} \bigg\{\bigg[\left(\frac{\tilde{\eta}(\omega_0)}{2m\omega^{\ddag}}\right)^2 +1\bigg]^{1/2}-\frac{\tilde{\eta}(\omega_0)}{2m\omega^{\ddag}} \bigg\},
\end{split}
\end{equation}
\nid where $k^\text{sys}_\text{TST}$ and $k_\text{TST}$ are the transition state rates for the system and system bath, respectively, $m$ the system mass, $\omega^{\ddag}$ the imaginary frequency of the system at the barrier top, and $\omega_0$ is given by the relation

\begin{equation}
\begin{split}\label{eq:Pollak}
\frac{\omega_0}{\omega^{\ddag}}  &=  \bigg[\left(\frac{\tilde{\eta}(\omega_0)}{2m\omega^{\ddag}}\right)^2 +1\bigg]^{1/2}-\frac{\tilde{\eta}(\omega_0)}{2m\omega^{\ddag}}.
\end{split}
\end{equation}

As elegantly proved by Pollak\cite{Pollak_JCP_1986}, Eq.~\ref{eq:Pollak} can be interpreted as a renormalized effective barrier frequency due to dissipation.
In a similar spirit, we would like to derive a relation between the RPI rates of the system without dissipation and the RPI-EF rates which include dissipation. Moreover, it would be desirable to obtain such a relation without resorting to any assumption on the potential energy surface.
The last condition forbids any direct relation between
$k_\text{inst}(
\beta,\eta)$ and $ 
k_\text{inst}(
\beta,\eta=0)$ at the same temperature, since both instanton pathways will have different extensions and therefore will be affected by different regions of the potential energy surface. Another possibility to tackle this problem is to ask the following question: \enquote{Given an instanton obtained at $T_a$ on $U_P^\text{sys}$, at which temperature $T_b$ will an instanton obtained on $U_P^\text{MF}$ present (approximately) the same geometry?}. Mathematically, given $\beta_a = 1/k_BT_a$ and $\bm{q}_a$ the solution to 

\begin{equation}
\begin{split}
0&=\frac{\partial S_P^\text{sys} (\beta_a,\bm{q}_a)}{\partial q_i^{(k)}}, \quad k=1,\dots, P  \quad i=1,\dots, 3N,
\end{split}
\end{equation}

\nid we aim to find $\beta_b$ such that $\bm{q}_a$ is an approximate solution to
\begin{equation}
\begin{split}
0&=\frac{\partial S_P^\text{MF} (\beta_b,\bm{q}_a)}{\partial q_i^{(k)}}, \quad k=1,\dots, P, \quad i=1,\dots, 3N.
\end{split}
\end{equation}

In the previous equations the sub-index $a$ and $b$ refer to the inverse temperatures $\beta_a$ and $\beta_b$, respectively.
Using Eq.~\ref{eq:S_P}
and going to the RP normal mode representation, we look for $
\beta_b$ that simultaneously satisfies
\begin{equation}
\begin{split}
0&=\frac{\partial S^\text{sys}(\beta_a,\textbf{q}_a)/\hbar}{\partial Q_i^{(l)}} = \beta^a_P\bigg[\frac{\partial V( \textbf{q}_a) }{\partial Q_i^{(l)}}
+ m_i{\omega_l^a}^2 Q_i^{(l)}
\bigg]\\
0&=\frac{\partial V( \textbf{q}_a)}{\partial Q_i^{(l)}}
+m_i{\omega_l^a}^2 Q_i^{(l)},
\end{split}\label{eq:cond1}
\end{equation}
\nid and
\begin{equation}
\begin{split}
0&=\frac{\partial S^\text{MF}(\beta_b,\textbf{q}_a)/\hbar}{\partial Q_i^{(l)}} = \beta^b_P\bigg[\frac{\partial V( \textbf{q}_b) }{\partial Q_i^{(l)}}
+ m_i{\omega_l^b}^2 Q_i^{(l)}
+\tilde{\eta}(\omega_l)Q_i^{(l)}
\bigg]\\
0&=\frac{\partial V( \textbf{q}_a)}{\partial Q_i^{(l)}}
+m_i{\omega_l^b}^2 Q_i^{(l)}
+\tilde{\eta}(\omega^b_l)\omega^b_l Q_i^{(l)},
\end{split}\label{eq:cond2}
\end{equation}
\nid for $l=-P/2+1,\dots, P/2$ and $i=1,\dots, 3N$. 

We combine Eq.~\ref{eq:cond1} and Eq.~\ref{eq:cond2} to get
\begin{equation}
\begin{split}\label{eq:Papa}
 {\omega_l^b}^2 
+\frac{\tilde{\eta}({\omega_l^b})}{m_i}{{\omega_l^b}}
-{\omega_l^a}^2
=0,
\end{split}
\end{equation}
\nid where $Q_i^{(l)}\ne 0$ was assumed since in that case Eq.~\ref{eq:cond1} and \ref{eq:cond2} became identical and  therefore the latter is trivially satisfied when the former is. 
In the limit of $P\to \infty$, we have $\omega_{l}=\frac{2\pi |l|}{\beta\hbar}$, so
we can solve the quadratic equation for $T_b$  to obtain 
\begin{equation}
\begin{split}\label{eq:Scaling}
 T_b/T_c^\circ = \sqrt{
 \bigg(\frac{ \tilde{\eta}(\omega_l^b)}{2m_i{\omega^\ddag}}\bigg)^2 \frac{1}{l^2}+ \left(\frac{T_a}{T_c^\circ}\right)^2
 } - 
 \frac{\tilde{\eta}(\omega_l^b)}{2m_i{\omega^\ddag}}\frac{1}{l},
\end{split}
\end{equation}
\nid where  $T_c^\circ$ is given by Eq.~\ref{eq:Tc}
and $l\neq0$. The previous equation cannot be fulfilled for all values of $l$ unless $\tilde{\eta}$ has a very specific frequency dependence. In part II of this paper, we will see that we can exploit the fact that the normal modes with $|l|=1$ dominate the instanton pathways, and thus we can solve Eq.~\ref{eq:Scaling} for $|l|=1$. Note that this equation has to be solved self-consistently, since $\omega^b_l$ depends on $T^b$. Interestingly, Eq.~\ref{eq:Tc-bath} results from Eq.~\ref{eq:Scaling} for $T_a=T_c$, so the latter equation can be interpreted as a generalization of the former. 
Eq.~\ref{eq:Scaling} allows one to compute the tunneling rates of a system coupled to a bath by only finding instanton pathways of the uncoupled system at a scaled temperature. Thus, it can be interpreted as a generalization of the GH equation into the deep tunneling regime.

\subsection{Position Dependent Friction}

The simplest system-bath coupling function that results in a position dependence of the friction is given by 

\begin{equation}
\begin{split}
f_{ij}(\b{q})= c_j g_i(\b{q}),
\end{split}\label{eq:separable_coupling}
\end{equation}
\nid leading to the following spectral density  
\begin{equation}
\begin{split}\label{eq:J-sep}
J_{il}(\b{q},\omega) =
\bigg(\frac{\partial g_{i}(\b{q})}{\partial q_l}\bigg)^2
\frac{\pi}{2}\sum_{j=1}^{N_b} \frac{c_j^2}{\mu_j\omega} (\delta(\omega-\omega_j)+\delta(\omega+\omega_j)).
\end{split}
\end{equation}

\nid This coupling function is equivalent to assuming 
that the zero-frequency value of the friction tensor $\tilde{\eta}(\bm{q},\omega=0)$ is position-dependent and its frequency dependence is identical for all positions~\cite{Straus_JCP_1993}. Thus, it is sometimes referred to as `separable coupling', and can be shown to yield a lower limit for the tunneling rate \cite{CALDEIRA_1983}. 
The MF-RP potential in this scenario becomes 
\begin{equation}
\begin{split}
U^{\text{MF}}_P
=& 
U^{\text{sys}}_P
+
\\&
\sum_{l=-P/2+1}^{P/2} \sum_{i=1}^{3N}  \sum_{j=1}^{N_b}  \frac{1}{2}\frac{c_j^2\omega_l^2}{\mu_j(\omega_j^2+\omega_l^2)\omega_j^2}
\bigg[ \sum_{k=1}^P C_{lk} g(\textbf{q}^{(k)})_i\bigg]^2\\
=&    
U^{\text{sys}}_P+
\\&\sum_{l=-P/2+1}^{P/2} \sum_{i=1}^{3N}  \sum_{j=1}^{N_b} 
\bigg[
\frac{1}{2}\frac{c_j^2\omega_l^2}{\mu_j(\omega_j^2+\omega_l^2)\omega_j^2}
\bigg] \times\\&
\bigg[ \sum_{k=1}^P C_{lk} \int_{0}^{1}ds\frac{d g(\textbf{q}^\text{k}(s))_i}{d s} 
- \sum_{k=1}^P C_{lk}g(\textbf{q}^\text{ref})_i
\bigg]^2\\
=&    
U^{\text{sys}}_P+
\\&\sum_{l=-P/2+1}^{P/2} \sum_{i=1}^{3N}  \sum_{j=1}^{N_b} 
\bigg[
\frac{1}{2}\frac{c_j^2\omega_l^2}{\mu_j(\omega_j^2+\omega_l^2)\omega_j^2}
\bigg]\times\\&
\bigg[ \sum_{k=1}^P C_{lk} \int_{0}^{1}ds\frac{d g(\textbf{q}^\text{k}(s))_i}{d s} 
\bigg]^2,\\
\end{split}
\end{equation}
\nid where we consider that $\textbf{q}^k(s):\mathbb{R} \rightarrow{} \mathbb{R}^{3N}$ is a parametrization, such that $\textbf{q}^k(0)=\textbf{q}^\text{ref}$ and
$\textbf{q}^k(1)=\textbf{q}^\text{k}$. In the last line we used that $\sum_{k=1}^P C_{lk}g(\textbf{q}^\text{ref})_i$ only contributes to the $l=0$ term and, since $\omega_0=0$, its contribution vanishes. As a consequence, the reference position is a free parameter which does not affect the results.

We continue by applying the chain rule
\begin{equation}
\begin{split}
U^{\text{MF}}_P=&    
U^{\text{sys}}_P
+\sum_{l=-P/2+1}^{P/2} \sum_{i=1}^{3N}  
\bigg[\sum_{k=1}^P C_{lk}  \int_0^{1}ds
 \sum_r \frac{\partial q^\text{k}_r (s)}{\partial s }\times\\&
\bigg(\sum_{j=1}^{N_b}
\frac{1}{2}\frac{c_j^2}{\mu_j \omega_j^2} 
\frac{\partial g(\textbf{q}^\text{k}(s))_i}{\partial q^\text{k}_r}^2
 \frac{\omega_l^2}{(\omega_j^2+\omega_l^2)}
 \bigg)^{1/2}
\bigg ]^2,
\end{split}
\end{equation}
and finally, by rearranging the terms, we obtain
\begin{equation}
\begin{split}\label{eq:eff-SD}
U^{\text{MF}}_P=&
U^{\text{sys}}_P +\\&
\sum_{l=-P/2+1}^{P/2}\sum_{i=1}^{3N}
\frac{\omega_l}{2}
\bigg[
\sum_{k=1}^P
C_{lk}
\bigg(
\sum_{r=1}^{3N}
\int_{\bm{q}^\text{ref}}^{\bm{q}^{(k)}} 
\tilde{\eta}_{ir}(\bm{q}',\omega_k)^{1/2}  \cdot d\bm{q_r}'\bigg)
\bigg]^2\\&
=U^{\text{sys}}_P +\\&
\sum_{l=-P/2+1}^{P/2}\sum_{i=1}^{3N}
\frac{\omega_l}{2}
\bigg[
\sum_{k=1}^P
C_{lk}
\bigg(\int_{\bm{q}^\text{ref}}^{\bm{q}^{(k)}} 
\tilde{\bm{\eta}_i}(\bm{q}',\omega_k)^{1/2}  \cdot d\bm{q}'\bigg)
\bigg]^2,
\end{split}
\end{equation}
\nid with $\tilde{\bm{\eta}}_i$ being the $i$-th row of the friction tensor.
We note that a straightforward extension of the Grote-Hynes approximation is not possible in this case. For completeness, for a one-dimensional system, the previous equation simplifies to 
\begin{equation}
\begin{split}
U^{\text{MF}}_P=& 
U^{\text{sys}}_P
+\sum_{l=-P/2+1}^{P/2} \frac{\omega_l}{2}
\bigg[\sum_{k=1}^P C_{lk}   \int_{q^\text{ref}}^{q^{(k)}} dq'  
 \tilde{\eta}(q',\omega_l)^{1/2} \bigg ]^2.
\end{split}
\end{equation}

\subsection{Renormalization of cross-over temperature}

    Naturally, the coupling of the bath to the system impacts the nuclear tunneling. One can study, for example, how the tunneling cross-over temperature is modified by the coupling to the bath. A trivial stationary point on the extended phase space of the ring polymer in the pathway that connects reactants and products can be found by locating all the beads at the top of the barrier. For a 1D system with position-dependent or independent friction under the   parabolic barrier approximation, one can write
\begin{equation}
\begin{split}
\lambda_l &=
 \sqrt{
 \omega_l^2 +
\frac{\tilde{\eta}(\omega_l)}{m} \omega_l -(\omega^\ddag)^2,
 } 
\end{split}
\end{equation}
\noindent where $\omega_l= 2\omega_P \sin(|l|\pi/P)$ are the free RP normal mode frequencies,  $i\omega^{\ddag}$ is the imaginary frequency at the barrier top, and $\tilde{\eta}$ has been evaluated at the barrier top.
In the  limit of large $P$, $\omega_l=2\pi|l|/\beta \hbar $ and the lowest three frequencies are
\begin{equation}
\begin{split}
\lambda_0 &= i\omega^{\ddag}\\
\lambda_{\pm 1} &=
 \sqrt{
 \frac{4\pi^2}{\beta^2\hbar^2} +
\frac{2\pi\tilde{\eta}(\omega_1)}{\beta \hbar m}  -(\omega^\ddag)^2
 } ,
\end{split}
\end{equation}
where $\omega_1$ refers to the first Matsubara frequency~\cite{Matsubara1955}, which depends on the temperature. 
The cross-over temperature is the temperature, $\beta^\text{sb}_c = 1/k_B T_c^\text{sb}$, below which 
$\lambda_{\pm 1}$ becomes imaginary (i.e. $\lambda_{\pm 1}^2$ becomes negative) and the location of the first-order saddle point is not at the top of the barrier, i.e. a non-trivial instanton pathway becomes possible. By taking $\lambda_{\pm 1}=0$ in the previous equation and solving the quadratic equation for $\beta^\text{sb}_c$, one obtains

\begin{equation}
\begin{split}
k_B T^\text{sb}_c & = \frac{1}{\beta^\text{sb}_c} = \frac{\hbar}{4\pi}\left(\sqrt{\frac{\tilde{\eta}(\omega_1)^2}{m^2}+4\omega^{\ddag 2}} - \sqrt{\frac{\tilde{\eta}(\omega_1)^2}{m^2}}\right),\label{eq:tc-bath-long}
\end{split}
\end{equation}
where $\omega_1$ is evaluated at $T_{c}^{\text{sb}}$.
Finally, identifying $T_c^\circ$ (Eq.~\ref{eq:Tc}) in the equation above leads to
\begin{equation}
\begin{split}
T_c^\text{sb}  = T^\circ_c \times 
\bigg[ 
\sqrt{ \bigg(\frac{\tilde{\eta}(\omega_1)}{2m \omega^\ddag}\bigg)^2 + 1} - \frac{\tilde{\eta}(\omega_1)}{2m \omega^\ddag}
\bigg].
\end{split}\label{eq:Tc-bath}
\end{equation}

Since $\omega_1$ depends on $T_c^\text{sb}$, Eq.~\ref{eq:Tc-bath} has to be solved self-consistently. The number between square brackets is always positive and less than 1, so $T_c^\text{sb}$ is always lower than $T_c^\circ$. Moreover, it is straightforward to see that the stronger the friction, the lower $T_c^\text{sb}$ becomes, and tunneling becomes less important at a given temperature. If the friction tensor is position-dependent,  $T_c^{\text{sb}}$ can be calculated by Eq.~\ref{eq:Tc-bath} replacing $\tilde{\eta}(\omega_1)$ by $\tilde{\eta}(\bm{q}^\ddag,\omega_1)$ where $\bm{q}^\ddag$ refers to the transition state geometry.

\section{\textit{Ab initio} Electronic Friction \label{sec:aifric}}

For systems in which the ground electronic state can be approximated by effectively independent quasi-particles such as the ones obtained with Kohn-Sham (KS) density-functional theory (DFT), the adiabatic electronic friction tensor  can be obtained from  first-principles simulations assuming non-interacting electrons and, as shown in Appendix C, adopts the following form for t>0

\begin{equation}
\begin{split}
\eta^\text{el}_{ij}(\textbf{q},t) =& 
\hbar \sum_{\nu,\nu'}
\braket{\psi_{\nu}|\partial_i \psi_{\nu'}}
\braket{\psi_{\nu'}|\partial_j\psi_{\nu}}
\Omega_{\nu \nu'}
\\
&\times
(f(\epsilon_\nu) - f(\epsilon_{\nu'}))
\cos{(\Omega_{\nu \nu'}t)},
\label{eq:eta_time}
\end{split}
\end{equation}
where $f(\epsilon)$ is the state occupation given by the Fermi-Dirac occupation function, $\Omega_{\nu\nu'}=(\epsilon_{\nu'}-\epsilon_{\nu})/\hbar$, 
$\psi_\nu$ and $\epsilon_\nu$ are the KS electronic orbitals  and orbital energies of the $\nu$-th level, $i$ and $j$ label the nuclear degrees of freedom, and $\partial_i = \partial/\partial q_i$.
A Fourier transform of the expression above 
leads to the usual expression employed in Refs. \cite{Head_Gordon_1995,Dou_PRL_2017,Maurer_2016_PRB} and reads 
\begin{equation}
\begin{split}\label{eq:gamma}
\hat{\eta}^\text{el}_{ij}( \textbf{q},\omega) =&
 \pi \hbar
\sum_{\nu, \nu'}
\braket{\psi_{\nu}| \partial_i\psi_{\nu'}}
\braket{\psi_{\nu'}|\partial_j\psi_\nu}
\times
(f(\epsilon_{\nu})-f(\epsilon_{\nu'}))
\\& 
\times
\Omega_{\nu\nu'}
 \delta (\Omega_{\nu{\nu'}} - \omega),
\end{split}
\end{equation}
where the k-point dependence has been omitted.






Most of the applications of the MDEF approach only consider an electronic friction tensor that is local in time to avoid the complexities of handling a non-instantaneous memory kernel and normally invoke the Markov approximation. This limit is also often referred to in the literature as the quasi-static limit since the Markov approximation is normally realized by taking $\omega \to 0$ limit in Eq.~\ref{eq:gamma}.\cite{Hellsing_1984}
In the cases where the system presents a constant density of states (DOS) around the Fermi level, an equivalent derivation is possible by applying the constant coupling approximation \cite{Head_Gordon_1995}.
The quasi-static limit involves the evaluation of the friction tensor in Eq.~\ref{eq:gamma} for excitations infinitesimally close to the Fermi level. 
In practical calculations with finite k-point grids, it is numerically challenging to accurately describe the DOS at the Fermi energy. This is typically circumvented by introducing a finite width for the delta function in Eq.~\ref{eq:gamma}. The choice of width depends on the system and, in literature, values between 0.01 and 0.60~eV can be found.\cite{Maurer_2016_PRB,Novko_PRL_2019, Connor_2021,Shipley_PhysRevB_2020}. 

 
The connection of the \textit{ab initio} electronic friction and  the RPI rate theory might seem a trivial substitution of Eq.~ \ref{eq:gamma} into Eq.~\ref{eq:eff-SD}. However, as we shall show in the next section, this is not the case and in order to obtain a better  connection between the electronic friction and the system-bath model used in the formulation of RPI-EF, a different expression should be employed.

\subsection{Electronic spectral density of non-interacting electrons}

Starting from Eq.~\ref{eq:eta_time}, and in a similar spirit to Eq.~\ref{eq:eta_t}, we perform a Laplace transform to get



\begin{equation}
\begin{split}
\tilde{\eta}_{ij}^\text{el}(\textbf{q},\lambda)&= \int_{0}^\infty dt e^{-\lambda t }\eta_{ij}^\text{el}(\textbf{q},t) \\ 
&=\hbar\sum_{\nu,\nu'}
\braket{\psi_{\nu} |\partial_i \psi_{\nu'}} 
\braket{\psi_{\nu'} | \partial_j \psi_{\nu}}
(f(\epsilon_\nu) - f(\epsilon_{\nu'}))\Omega_{\nu\nu'}\\
&\times
\frac{\lambda}{\lambda^2 + \Omega_{\nu\nu'}^2}.
\label{eq:eta6}
\end{split}
\end{equation}

The equation above adopts the same limit for $\lambda \to 0$ as Eq.~\ref{eq:gamma}. However, for $\lambda > 0$, instead of the $\delta$ function, we obtain a sum of Lorentzian functions of width $2 \lambda$. Comparing Eq.~\ref{eq:eta6} and \ref{eq:eta_laplace}, we can identify the equivalent of the spectral density in RPI-EF as

\begin{equation}
\begin{split}\label{eq:friction_spectral_density}
    J_{ij}(\bm{q}, \omega) = \pi \hbar \sum_{\nu, \nu'}
\braket{\psi_\nu | \partial_i \psi_{\nu'}} \braket{\psi_{\nu'} | \partial_j \psi_\nu} \omega^2 \\
 \times (f(\epsilon_\nu) - f(\epsilon_{\nu'}))\delta(\omega-\Omega_{\nu\nu'}),
\end{split}
\end{equation}

\nid which provides a seamless connection between RPI-EF and electronic friction.  

In RPI-EF, the spectral density of electronic friction shown in Eq.~\ref{eq:friction_spectral_density} is evaluated simultaneously at the ring polymer normal mode frequencies. Thus, we have derived viable expressions to combine RPI-EF with an electronic friction formulation that can be calculated from first-principles, without any further approximations, except for the assumption of separable coupling, which we shall examine for real systems in part II of this paper. We expect that the connection of these two theories will be suitable as long as the approximations of both underlying theories remain valid.

\section{ Conclusions }\label{sec:Conclusions}

We have presented an extension of the ring-polymer instanton rate theory to describe a system coupled to a bath of harmonic oscillators through the definition of an effective friction tensor that enters the instanton ring-polymer potential energy expression. We therefore refer to this method as the RPI-EF approach.
The theory is rather general and allows the inclusion of frequency and position dependence in the system-bath coupling for the calculation of thermal tunneling rates within the instanton approximation.
For the case of linear coupling, we derived an approximation that allows one to predict RPI-EF reactions rates using only RPI calculations. The approximation can be understood as an extension of the Grote-Hynes approximation to the deep tunneling regime. This may be useful to estimate whether it is necessary to carry out full RPI-EF calculations for a particular reaction. 

RPI-EF is a method tailored for the description of tunneling rates and based on imaginary-time trajectories. Therefore, it cannot be applied for the simulation of vibrational relaxation or scattering experiments \cite{Bunermann_Sci_2015, Wodtke_ChemSocRev_2016, JiangMillerBunermann2021}, where some kinds of NQEs and NAEs could interplay strongly. It would be interesting to write similar extensions to approaches based on path integral molecular dynamics~\cite{Craig_JCP_2004, CMD2, Rossi_JCP_2014, Trenins_JCP_2019}, since they would yield efficient approximations to model these situations. 
Other future directions could cover the inclusion of non-equilibirum effects through the use of non-positive definite friction tensors~\cite{Bode_2012,Brandbyge_SurfSci_2019}.
We hope that the derivations presented in this work stimulate further theoretical developments in this area and allow new phenomena to be explained in situations that we have not yet explored. 

\begin{acknowledgments}
Y.L., E.S.P. and M.R. acknowledge financing from the Max Planck Society and computer time from the Max Planck Computing and Data
Facility (MPCDF). Y.L and M.R. thank Jeremy Richardson, Aaron Kelly, and Stuart Althorpe for a critical reading of the manuscript.
C.L.B. acknowledges financial support through an EPSRC-funded PhD studentship. R.M. acknowledges Unimi for granting computer time at the CINECA HPC center. R.J.M. acknowledges financial support through a Leverhulme Trust Research Project Grant (RPG-2019-078) and the UKRI Future Leaders Fellowship programme (MR/S016023/1).

\end{acknowledgments}

\pagebreak

\section*{Appendix}

\subsection{Free Ring Polymer Normal Modes \label{sec:NM}}

The free ring polymer potential is given by
 setting $V=0$ in Eq.~\ref{eq:U_P}. The resulting potential is harmonic, however, due to the presence of degenerate eigenvalues there is no unique transformation to diagonalize it. Assuming $P$ is even, one  possibility is the following orthogonal coordinate transformation\cite{Craig_thesis, Markland_JCP_2008}

\begin{equation}
Q^{(l)}_{i} = \sum^P_{k=1} C_{lk}^{(P)}{q}^{k}_{i}  \normalfont{ \quad    i=1, \dots, 3N, \quad l=-P/2+1, \dots ,P/2},\label{eq:C-NM}
\end{equation}
\nid where the  $P \times P$  matrix $C^{(P)}$ is defined as
\begin{gather*}\label{eq:matrix_C2}
C_{lk}^{(P)} =
\begin{cases}
  \sqrt{\frac{1}{P}}  & l = 0 \\
  \sqrt{\frac{2}{P}} \cos(\frac{2 \pi kl}{n}) & 1\le l \le P/2-1\\
\sqrt{\frac{1}{P}} (-1)^j & l = P/2  \\
  \sqrt{\frac{2}{P}} \sin(\frac{2 \pi kl}{P}) & -P/2+1\le l\le -1.\\
\end{cases}
\end{gather*}

\subsection{Mean-Field Ring Polymer Potential in Cartesian Representation for Spatially Independent Coupling}

The mean-field RP potential in Cartesian representation for the linear coupling case is obtained by introducing Eq.~\ref{eq:C-NM} into \ref{eq:MF-SI} leading to 
\begin{equation}
\begin{split}
U^{\text{MF}}_P =&
\sum_{l=1}^P V(q_1^{(l)},\dots, q_{3N}^{(l)})
          +\\&
 \sum_{\substack{k=1\\k'=1}}^P 
 \sum_{i=1}^{3N} 
   \frac{1}{2}q_i^{(k')}O_{i,k,k'} q_i^{(k)}
+
   \frac{1}{2}q_i^{(k')}D_{i,k,k'} q_i^{(k)}
\end{split}
\end{equation}
\nid with

\begin{equation}
\begin{split}
O_{k,k'} = m_i\omega_P^2 (2\delta_{k,k'}-\delta_{k,k'-1} - \delta_{k,k'+1})
\end{split}
\end{equation}

\nid and 
\begin{equation}
\begin{split}
D_{k,k'} =& \eta_0\omega_P\bigg[ \sum_{l=1}^{(P-1)/2}\frac{4}{P}\sin(\pi l /P)\cos(2\pi l (k-k')/P) 
\\& - \frac{2}{P}(-1)^{k+k'}\bigg],
\end{split}
\end{equation}

\begin{figure}[ht]
\centering
       \includegraphics[width=1.\columnwidth]{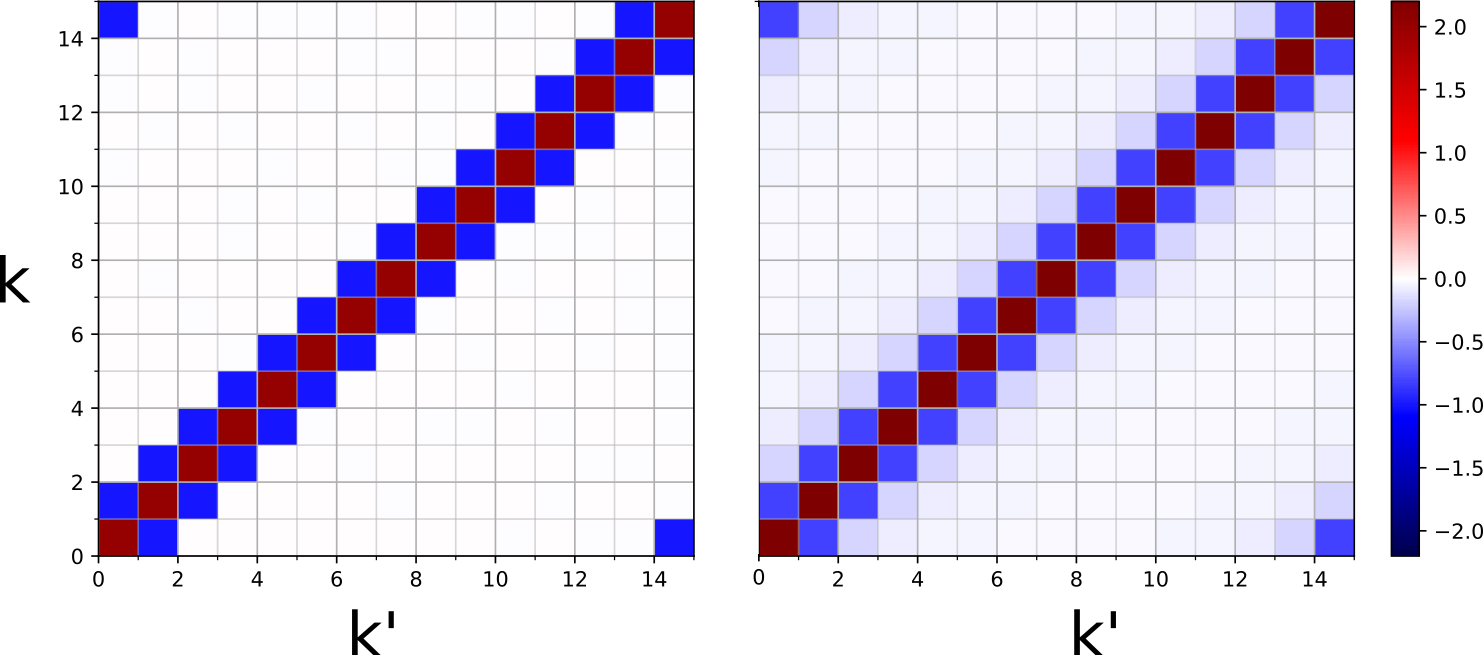}
    \caption{ Spring coupling matrices $O_{k,k'}$ (left) and $D_{k,k'}$ (right)     at 25 K for  15 beads and $\eta_0/m\omega_0=1.0$. The values have been scaled by $m\omega_P$ to ease visual comparison. } 
    \label{fig:NM-Coeff}
\end{figure}

\nid where  a linear (Ohmic) spectral density was considered.
In Fig. \ref{fig:NM-Coeff} we show a graphical representation of the spring coupling matrices $O_{k,k'}$ and  $D_{k,k'}$. The latter has small but non-zero matrix elements outside the tridiagonal entries present in the former, representing coupling beyond nearest-neighbor beads. Moreover, the matrix elements decay rapidly with the increase of the bead index distances when periodic boundary conditions are considered.

\subsection{Arriving at Eq.~\ref{eq:eta_time}}\label{sec:DMS}

The steps presented in this Appendix closely follow Ref. \cite{Dou_PRL_2017}, and we repeat them here merely for completeness.
We consider a quadratic electronic Hamiltonian of the form,
\begin{equation}
\begin{split}
\hat{h} = \sum_{pq}h_{qp}(\mathbf{q})\hat{d}_p^+\hat{d}_q
\end{split}
\end{equation}
\nid where $h_{qp}(\textbf{q})$ is general notation to represent that matrix elements might depend on the nuclear degrees of freedom, $\textbf{q}$, and $\hat{d}_p^+$ and $\hat{d}_q$ are the electronic creation and annihilation operators, respectively.

Starting from the quantum-classical Liouville equation, the electronic friction tensor in the adiabatic limit  with nuclei fixed at position $\textbf{q}$ can be written as \cite{Dou_PRL_2017}

\begin{equation}
\begin{split}
\eta^\text{el}_{ij}(\textbf{q},t) =-\text{tr}_e\bigg( \partial_i \hat{h}e^{-i\hat{\hat{\mathcal{L}}} t} \partial_j \hat{\rho}_{ss}\bigg) \label{eq:eta-dou}
\end{split}
\end{equation}
\nid where $\hat{\hat{\mathcal{L}}}$ is the Liouvillian superoperator, $\text{tr}_e$ implies tracing over the electronic degrees of freedom, $\hat{\rho}_{ss}$ is the steady state electronic density matrix, and $i$ and $j$ represent two nuclear degrees of freedom.

By recalling that $e^{-i\hat{\hat{\mathcal{L}t}}}(\cdot) =e^{-i\hat{h}t/\hbar}(\cdot)e^{i\hat{h}t/\hbar} $, the invariance of the trace under cyclic permutations, and considering the quadratic Hamiltonian presented above we find
\begin{equation}
\begin{split}
\eta^\text{el}_{ij}(\textbf{q},t) &=-\text{tr}_e\bigg( e^{i\hat{h}t}\partial_i \hat{h}e^{-i\hat{h}t} \partial_j \hat{\rho}_{ss}\bigg)\\
&= -\sum_{mn}\partial_i h_{nm}
\text{tr}_e\bigg( e^{i\hat{h}t}\hat{d}_m^+\hat{d}_n e^{-i\hat{h}t} \partial_j \hat{\rho}_{ss}\bigg)\\
&= -\sum_{mn}\partial_i h_{nm}
\text{tr}_e\bigg( e^{i\hat{h}t}\hat{d}_m^+e^{-i\hat{h}t}e^{i\hat{h}t}\hat{d}_n e^{-i\hat{h}t} \partial_j \hat{\rho}_{ss}\bigg).\\
\label{eq:eta2}
\end{split}
\end{equation}

By noting that $ e^{i\hat{h}t/\hbar}\hat{d}_m^+e^{-i\hat{h}t/\hbar} = \sum_a (e^{i \hat{h} t/\hbar})_{ma}\hat{d}_a^+$, and 
 $ e^{i\hat{h}t/\hbar}\hat{d}_n e^{-i\hat{h}t/\hbar} = \sum_b (e^{-i \hat{h} t/\hbar})_{bn}\hat{d}_b$ (see Supporting Information in Ref. \cite{Dou_PRL_2017}),  and defining $\sigma_{ab}^{ss} = \text{tr}_e(\hat{d}_a^+\hat{d}_b \hat{\rho}_{ss})$, Eq. \ref{eq:eta2} can be expressed as 
\begin{equation}
\begin{split}
\eta^\text{el}_{ij}(\textbf{q},t) &=
-\sum_{mnab}\partial_i h_{nm}
(e^{i\hat{h}t/\hbar})_{ma} (e^{-i\hat{h}t/\hbar})_{bn}
\text{tr}_e\bigg( \hat{d}_a^+\hat{d}_b \partial_j \hat{\rho}_{ss}\bigg)\\
&=
-\sum_{mnab}\partial_i h_{nm}
(e^{i\hat{h}t/\hbar})_{ma}
\partial_j \sigma_{ab}^{ss}
(e^{-i\hat{h}t/\hbar})_{bn}
\\
&=
-\sum_{n}\bigg(\partial_i \hat{h}
e^{i\hat{h}t/\hbar}
\partial_j\sigma^{ss}
e^{-i\hat{h}t/\hbar}
\bigg)_{nn}
\label{eq:eta3}
\end{split}
\end{equation}
\nid where $\sum_{n}$ represents a sum over electronic orbitals.
If we take the basis in which $\hat{h}$ is diagonal, 
\begin{equation}
\begin{split}
\hat{h} = \sum_\nu \epsilon_\nu  \Ket{\psi_{\nu}}\Bra{\psi_{\nu}},
\end{split}\label{eq:h_kk}
\end{equation}

\nid Eq.~\ref{eq:eta3} simplifies to
\begin{equation}
\begin{split}
\eta^\text{el}_{ij}(\textbf{q},t) &=
-\sum_{\nu,\nu'}
\Bra{\psi_{\nu}}\partial_i \hat{h}\Ket{\psi_{\nu'}}
e^{i\epsilon_{\nu'}t/\hbar}
\Bra{\psi_{\nu'}}\partial_j \sigma^{ss}\Ket{\psi_{\nu}}
e^{-i\epsilon_{\nu} t/\hbar}.
\label{eq:eta4}
\end{split}
\end{equation}

At equilibrium, we can write $\hat{\sigma}^{ss}$ as 
\begin{equation}
\begin{split}
\hat{\sigma}^{ss} = \sum_\nu f(\epsilon_\nu) \Ket{\psi_{\nu}}\Bra{\psi_{\nu}},
\end{split}\label{eq:sigma_ss}
\end{equation}

\nid and therefore, the second matrix element in Eq.~\ref{eq:eta4} can be evaluated as
\begin{equation}
\begin{split}
\Bra{\psi_{\nu'}} \partial_i\hat{\sigma}^{ss}\Ket{\psi_{\nu}}
 =& 
\partial_i \epsilon_\nu \frac{f(\epsilon_k)}{\partial \epsilon_\nu}\delta_{\nu \nu'}  
\\&+(f(\epsilon_\nu) - f(\epsilon_{\nu'}))
\braket{\psi_{\nu'} |\partial_i\psi_\nu} \label{eq:partial_nu}.
\end{split}
\end{equation}

Introducing Eq.~\ref{eq:partial_nu} into Eq.~\ref{eq:eta4},
and using 
\begin{equation}
\begin{split}
  \partial_i   \Bra{\psi_\nu} \hat{h}\Ket{\psi_{\nu'}} &= \partial_i  \epsilon_\nu \delta_{\nu\nu'}\\&= \Bra{\psi_\nu}\partial_i \hat{h}\Ket{\psi_{\nu'}} +\braket{\psi_\nu |\partial_i\psi_{\nu'}} (\epsilon_\nu - \epsilon_\nu'), 
\end{split}
\end{equation}

\nid leads to  

\begin{equation}
\begin{split}
\eta^\text{el}_{ij}(\textbf{q},t) =& 
\sum_{\nu,\nu'}
\Bigg[
\braket{\psi_{\nu}|\partial_i \psi_{\nu'}}(\epsilon_\nu' - \epsilon_\nu) +
\partial_i \epsilon_\nu \delta_{\nu\nu'}
\Bigg]\\
&\times
\Bigg[
\partial_j \epsilon_\nu \frac{f(\epsilon_\nu)}{\partial \epsilon_\nu}\delta_{\nu\nu'} + (f(\epsilon_\nu) - f(\epsilon_\nu'))
\braket{\psi_{\nu'}|\partial_j\psi_{\nu}}
\Bigg]\\
&\times
e^{-i(\epsilon_\nu -\epsilon_{\nu'})t/\hbar}.
\label{eq:6}
\end{split}
\end{equation}

Finally, upon noticing that the factors in front of the exponential are invariant under the exchange of orbital labels such that the imaginary part of the complex exponential vanishes,
and the fact that for $\nu=\nu'$ the expression is zero,
one arrives to Eq. \ref{eq:eta_time} presented in the main text. This expression agrees
with an alternative recent derivation based on the exact factorization of the electronic-nuclear wavefunction~\cite{Martinazzo_2021}.



%
\bibliographystyle{aipnum4-1}

\end{document}